\documentclass[twoside]{article}
\usepackage{fleqn,espcrc2}
\usepackage{graphicx}

\def\cascade{{\sc Cascade}}
\def\herwig{{\sc Herwig}}
\def\pythia{{\sc Pythia}}
\def\mcatnlo{MC@NLO}
\def\prp{\perp}
\def\kt{\ensuremath{k_\prp}}

\newcommand{\AmS}{{\protect\the\textfont2
  A\kern-.1667em\lower.5ex\hbox{M}\kern-.125emS}}

\begin{document}

\begin{figure}[htb]
\includegraphics{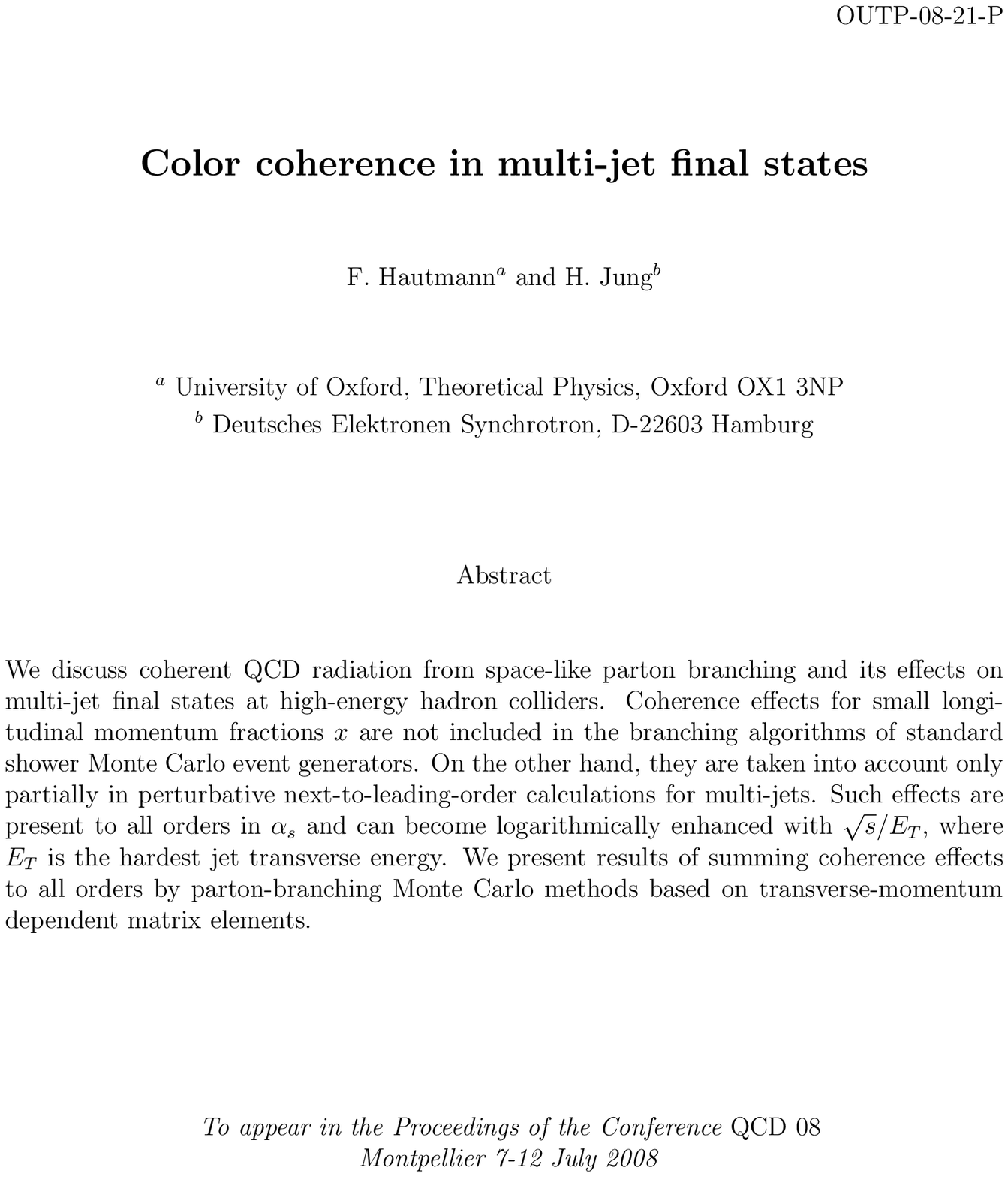}
\end{figure}

\title{Color coherence   in   multi-jet final states}

\author{F. Hautmann\address{University of Oxford, Theoretical Physics, 
        Oxford OX1 3NP}
        and 
        H. Jung\address{Deutsches Elektronen Synchrotron, 
       D-22603 Hamburg }}

\begin{abstract}
We discuss    coherent  QCD radiation   from  space-like parton branching    
and its effects   on multi-jet final states   at high-energy  hadron colliders. 
   Coherence effects  for small 
   longitudinal momentum fractions $x$    
   are not included in the branching  algorithms of standard 
shower    Monte Carlo  event generators. On the  other hand, they are 
  taken into account  only partially in perturbative 
  next-to-leading-order   calculations for  multi-jets. 
  Such effects are present to all orders 
  in $\alpha_s$ and can  become  logarithmically enhanced  with $\sqrt{s} / E_T$, where 
  $E_T$ is the hardest jet  transverse energy.   We present results of summing 
  coherence effects to all orders by parton-branching Monte Carlo methods based  on 
  transverse-momentum dependent  matrix elements. 
\end{abstract}

\maketitle

Studies of   final states  containing  multiple hadron jets   at the forthcoming 
LHC require realistic   event simulation  
by parton shower Monte Carlo~\cite{mlmhoche}, including 
validation and tuning   of the event generators 
using  multi-jet data from Tevatron and HERA,  
see e.g.~\cite{albrow,d02005,zeus1931}.  

Standard parton-shower generators, such as 
\herwig\ and \pythia,   are based on 
collinear evolution of the jets developing  (both ``forwards" and ``backwards") 
 from the hard event.     Besides 
small-angle, incoherent  gluon emission,     
the approach of these generators  also  
 incorporates  
  coherent   soft-gluon emission from partonic lines carrying longitudinal 
momentum fraction $x \sim {\cal O} (1)$. The  phenomenological relevance 
  of this  treatment  has been emphasized by  extensive   collider data  
  studies~\cite{mclectures}. 

\begin{figure}[htb]
\vspace{40mm}
\includegraphics{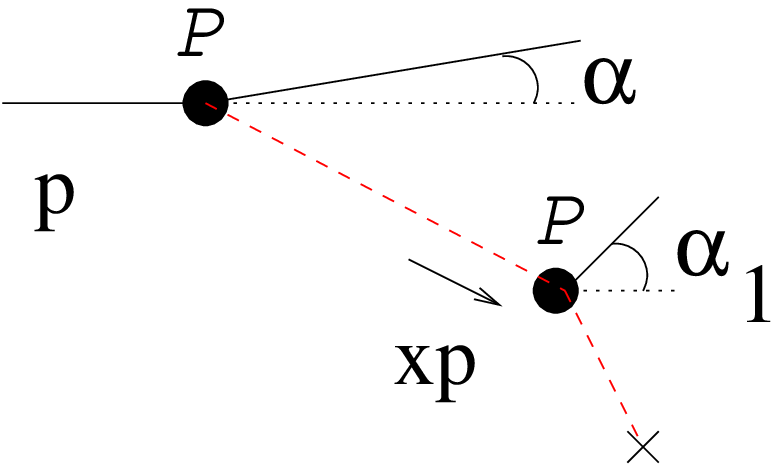}
\includegraphics{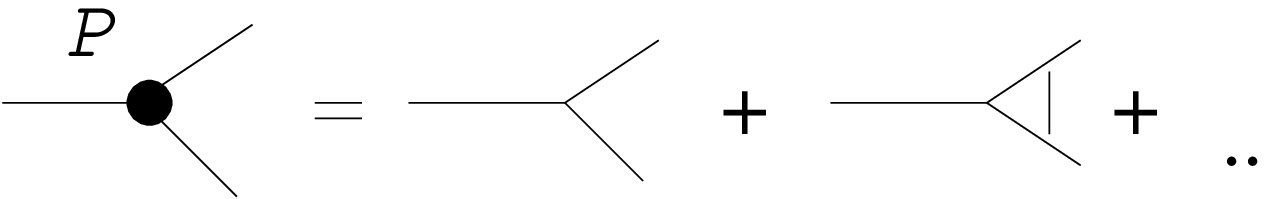}
\caption{(top) Coherent radiation 
 in the space-like parton shower for $x \ll 1$; (bottom) the unintegrated 
splitting function ${\cal P}$, including small-$x$ virtual 
corrections.} 
\label{fig:coh} 
\end{figure}

 However  
 at the LHC,   due to the phase space opening up for large center-of-mass 
 energies,   
 jet production enters a  new regime    with a great many 
   events  characterized by  multiple hard scales, 
  in which 
 (a) effects of emissions that are not collinearly ordered 
 become increasingly non-negligible,  and (b) coherence effects set in 
 from space-like partons carrying momentum fractions $x \ll 1$ 
 (Fig.~\ref{fig:coh})~\cite{hj_ang}. 
These effects are not included in standard shower Monte Carlo generators.  
It is not at all 
obvious that the approximations involved in 
Monte Carlo that have successfully served for event 
simulation in past collider experiments will be up to the new situation.

The theoretical framework to take  account of non-collinear emission 
and coherence in the space-like branching 
requires the introduction  of partonic distributions 
unintegrated not only in the longitudinal momenta 
but also in the  transverse momenta~\cite{skewang,hef,mw,anderss96,hj_rec}. 
The   corrections to collinear ordering  
correspond  to  higher-order radiative 
 terms in  the associated 
jet distributions  that are  logarithmically enhanced in the ratio  $\sqrt{s} / E_T$  of 
the total energy $\sqrt{s}$ to the   jet transverse energy. 
An alternative approach to non-collinear gluon radiation 
is based on  showers of color dipoles~\cite{gustaf86}  
and  is also being applied to the initial-state jet~\cite{lonn,gustaf08}.  
 See~\cite{marchedok}   for a study 
 of  critical  issues in the  relation of this approach with  the parton  formulation. 
 Either  at parton  or  dipole level,  open questions 
involve methods for  properly combining 
contributions from infrared regions  with high-energy subgraphs,
for which  we expect techniques  
such as those in~\cite{jccfh}  to  be relevant. 
An overview of  issues on  unintegrated distributions 
and references may be found in~\cite{hj_rec}. 

\begin{figure}[htb]
\vspace{44mm}
\includegraphics{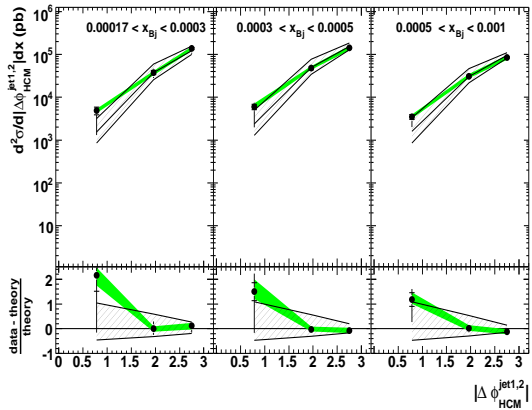}
\caption{Measurement of 
3-jet   cross section~\protect\cite{zeus1931} 
  versus azimuthal separation between the two 
highest-$E_T$ jets, compared with 
NLO results.} 
\label{fig:phizeus-3j}
\end{figure}

The effects  of   coherent  space-like branching 
are investigated   in~\cite{hj_ang} 
 for   angular and momentum 
correlations  in  multi-jet  final states.   
 For a multi-jet event, 
consider for instance the  distribution in the azimuthal angle 
$\Delta \phi$  between the two hardest jets. 
 At the LHC such  measurements  
  may  become  accessible    relatively early 
  and   be used to probe   the 
  description  of  complex hadronic  final  states by QCD and Monte Carlo generators. 
Experimental data on 
     $\Delta \phi$  correlations are available  from the 
    Tevatron~\cite{d02005}  and from Hera~\cite{zeus1931}. 
  While the Tevatron  measurements are  dominated~\cite{hj_ang}  by 
    leading-order QCD processes and are reasonably well described both 
  by  collinear showers  (\herwig\ and  
  the new tuning of \pythia~\cite{albrow,d02005}) and  by 
  fixed-order NLO 
  calculations, 
   the     Hera   $\Delta \phi$  measurements  probe 
   higher orders in the dynamics of color emission, and  present a more complex 
case,  
likely to be  closer  to the situation 
at the LHC. 

\begin{figure}[htb]
\vspace{114mm}
\includegraphics{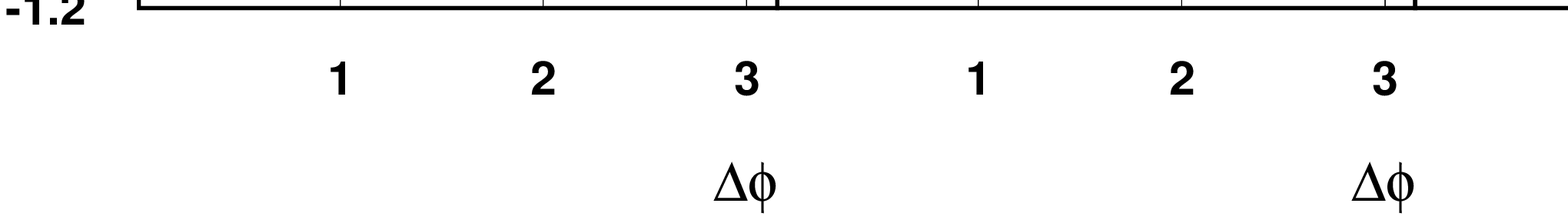}
\includegraphics{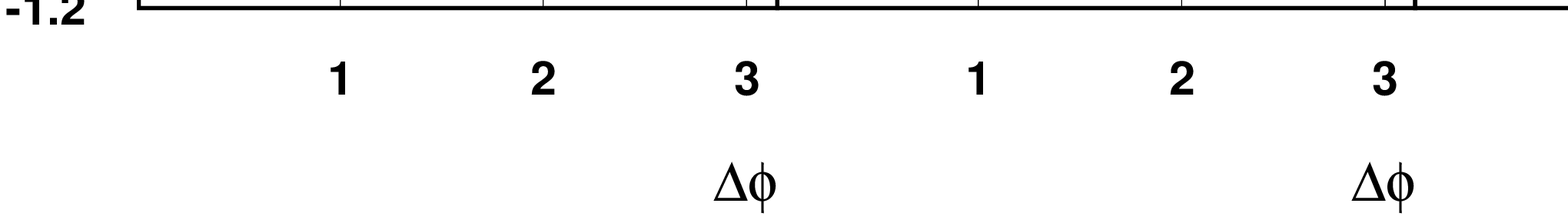}
\caption{Angular jet correlations~\protect\cite{hj_ang} 
from \kt-shower 
\cascade\ and  from  \herwig, compared 
with   data~\protect\cite{zeus1931}: 
(top) di-jet cross section; (bottom) three-jet cross section. 
The \herwig\    results are multiplied by a K-factor equal to 2 to give the correct 
normalization in the two-jet region. 
} 
\label{fig:phipage}
\end{figure}

In particular, it is  noted   in~\cite{hj_ang,hjradcor}  that  
di-jet $\Delta \phi$ correlations are affected by sizeable  
sub-leading corrections,  resulting in large theoretical uncertainties at NLO. 
Analogous effects are observed~\cite{zeus1931} in the three-jet cross section
(Fig.~\ref{fig:phizeus-3j}~\cite{zeus1931}) 
particularly for the  small-$\Delta \phi$ and small-$x$  bins. 
The large corrections arise 
from regions 
with three well-separated hard jets in which 
 the parton  lines in the initial state decay chain 
 are not ordered in transverse momentum.  
 These corrections can be treated and summed to all  orders, including 
 coherence effects,  
 by parton branching~\cite{hj_ang},  
 using  matrix elements (ME) and 
 distributions (u-pdf) at fixed   transverse 
 momentum  k$_\perp$  according to the factorization~\cite{hef}. 
 Fig.~\ref{fig:phipage}  compares k$_\perp$-shower (\cascade~\cite{jung02}) and 
 collinear-shower 
  (\herwig)  results with the measurements~\cite{zeus1931}.   
  The shape of the angular distribution is 
  described well by the k$_\perp$-shower. 
On the  
other hand, while the \herwig\ K-factor in  Fig.~\ref{fig:phipage}  allows  one to get 
the normalization in 
the two-jet region approximately correct, the shape is not well described by 
\herwig\ as  $\Delta \phi$ decreases.

\begin{figure}[htb]
\vspace{40mm}
\includegraphics{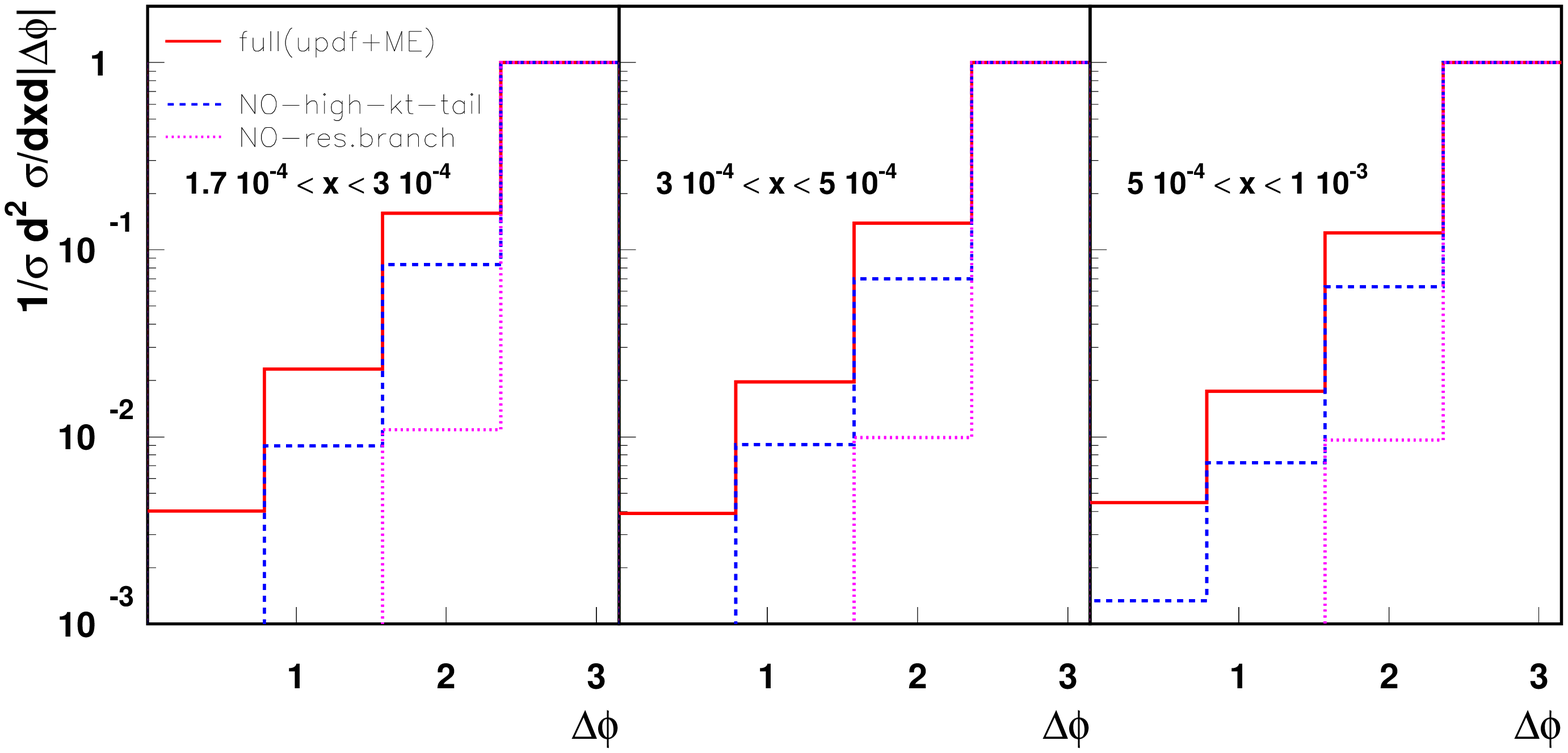}
\caption{The dijet azimuthal distribution~\protect\cite{hj_ang} 
    normalized to the
back-to-back cross section:
(solid red)  full result
(u-pdf $\oplus$ ME); (dashed blue) no finite-\kt
correction in ME
 (u-pdf $\oplus$ ME$_{collin.}$);
(dotted violet) u-pdf with no resolved branching.
}
\label{fig:ktord}
\end{figure}

The   correlations in the jets'   azimuthal distances and in their transverse-momentum 
imbalance are sensitive to the 
large-k$_\perp$ tail in the hard matrix element~\cite{ch94}. 
Fig.~\ref{fig:ktord} illustrates  the relative contribution of ME and u-pdf 
 to   the   result, 
showing different approximations to the
azimuthal dijet distribution normalized to the
back-to-back cross section. The solid
red curve is   the full  result. 
  The dashed blue curve is  obtained
from the same unintegrated pdf's but
by taking the collinear approximation in
the hard matrix element.    The dashed curve  
drops much faster than the full result as $\Delta \phi$ decreases, 
indicating  that  the
high-\kt component  in the hard ME 
is necessary  to describe 
jet correlations      
for small $\Delta \phi$. 
The dotted (violet) curve is  the result 
  obtained from the 
unintegrated pdf 
without any resolved branching.  
 This   represents   the contribution of 
the intrinsic \kt  distribution only, 
corresponding
 to nonperturbative, predominantly
 low-\kt modes.  That is,  in the dotted (violet) curve one retains 
an  intrinsic 
 \kt $\neq 0$ but no  effects of coherence. We see  
 that the resulting jet correlations in this case are down by an order of magnitude. 
The inclusion of  the  perturbatively computed  high-k$_\perp$ 
 correction    
 distinguishes the  calculation~\cite{hj_ang}  of  multi-jet cross sections 
 from other  shower approaches (see e.g.~\cite{hoeche})  
 that include transverse momentum 
dependence in the  pdfs but not  in the  matrix elements.

\begin{figure}[htb]
\vspace*{4.8 cm}
\includegraphics{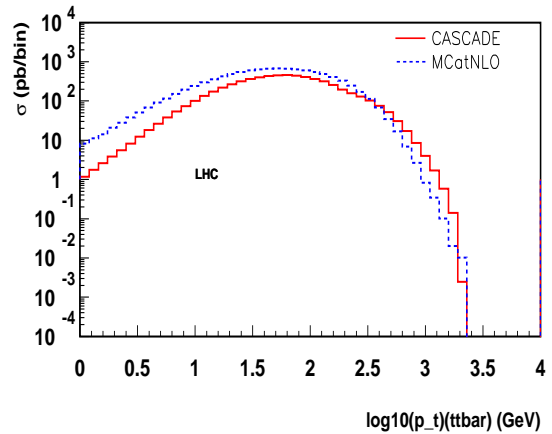}
\caption{Transverse momentum 
distribution of top quark  pairs~\protect\cite{hj_ang}  at the LHC  from
 k$_\perp$-shower  (\cascade)  
and    NLO+shower   (\mcatnlo).} 
\label{fig:ttbar}
\end{figure}

We  note that this behavior  in the finite-k$_\perp$  part of the matrix element  
is consistent with the  observation 
  of  sizeable  corrections to the  $\Delta \phi$ 
distribution at NLO~\cite{zeus1931}, and supports the interpretation 
of the data in   Fig.~\ref{fig:phipage}  
in terms of corrections to collinear ordering. 

 The  coherence effects just discussed  show up in the 
region of small $\Delta \phi$.  The region of large $\Delta \phi$, on the other hand, 
is dominated by  soft and collinear radiation~\cite{delenda}.   
 But note that this region  is also possibly affected by further, somewhat unexpected 
 effects,  arising from endpoint singularities~\cite{endp}, 
and possibly   Coulomb exchange~\cite{manch}.

Besides jet final states,   
the corrections to collinear-ordered showers 
 described in this article are  also relevant to 
 heavy particle production~\cite{hef,hgsetc,deak,wz}. 
In   Fig.~\ref{fig:ttbar}  is the result of 
 a numerical calculation for the transverse momentum spectrum of 
 top-antitop pair production at the   LHC. Small-$x$ effects are not large in this case. 
Rather,  this process  illustrates  how the  showering  works   all the way up to   
 virtualities  on   the order of the top quark mass, and finite $x$.   
 Note that even at LHC energies the transverse momentum distribution 
of top quark pairs calculated from the k$_\perp$-shower is similar to what is obtained 
from a full 
NLO calculation (including parton showers, MC@NLO),  
with the  k$_\perp$-shower  giving a somewhat harder spectrum.

\end{document}